\documentclass[conference,10pt]{IEEEtran}
\IEEEoverridecommandlockouts

\usepackage{epsfig,rotating,setspace,latexsym,amsmath,epsf,amssymb,amsfonts,bm,theorem,cite,algorithm,graphicx,epsf,authblk,epstopdf,color,algpseudocode,bbm}

\usepackage[algo2e]{algorithm2e}

\newtheorem{remark}{Remark}

\newtheorem{lemma}{Lemma}
\newenvironment{Proof}[1]{\medskip\par\noindent{\bf Proof:\,}\,#1}{{\mbox{\,$\blacksquare$}\par}}

\allowdisplaybreaks

\begin{document}

\title{Private Status Updating with Erasures:\\A Case for Retransmission Without Resampling\thanks{This work was supported by the U.S. National Science Foundation under Grants CNS 21-14537 and ECCS 21-46099.}
}

\author[1]{Ahmed Arafa}
\author[2]{Karim Banawan}
\affil[1]{\normalsize Electrical and Computer Engineering Department, University of North Carolina at Charlotte, USA}
\affil[2]{\normalsize Department of Electrical Engineering, Alexandria University, Egypt}

\maketitle

\begin{abstract}
A status updating system is considered in which a source updates a destination over an erasure channel. The utility of the updates is measured through a function of their {\it age-of-information} (AoI), which assesses their freshness. Correlated with the status updates is another process that needs to be kept {\it private} from the destination. Privacy is measured through a {\it leakage} function that depends on the amount and time of the status updates received: stale updates are more private than fresh ones. Different from most of the current AoI literature, a {\it post-sampling waiting} time is introduced in order to provide a privacy cover at the expense of AoI. More importantly, it is also shown that, depending on the leakage budget and the channel statistics, it can be useful to retransmit stale status updates following erasure events {\it without resampling} fresh ones.
\end{abstract}

\section{Introduction}

Providing fresh status updates to destinations is crucial for timely decision-making in various applications, including smart city, e-Health, and digital twins, to name a few. At the same time, with the vast connectivity and correlation between data sources, some information may need to be kept private from curious destinations while providing them with the useful data they need. In this paper, we consider the interplay between utility and privacy of status updates through the lens of {\it time.}

Data freshness is quantified by the age-of-information (AoI), defined as the time elapsed since the latest useful piece of received data has been generated \cite{yates2021age}. In this paper, the utility of status updates is measured through a function acting on their AoI. Such function may represent the estimation mean square error \cite{sun-wiener, arafa-banawan-seddik-sample}, under some assumptions on the underlying process being updated. Privacy, on the other hand, is measured through a function that depends on the relationship between the amount of data received so far and the process to be kept private. We focus on scenarios in which the {\it privacy leakage is at its peak when status updates are most fresh.} Thereby, a tension arises between data freshness and data privacy.

We study a continuous-time status updating system in which a source-destination pair are communicating through an erasure channel. The freshness of data is controlled by {\it pre-sampling} waiting times \cite{sun-age-mdp}, while the privacy is maintained by {\it post-sampling} waiting times. The post-sampling waiting times are carefully designed to deliver {\it moderately fresh} updates; these are updates that are fresh enough to provide utility, yet stale enough to provide privacy. A main pillar in our work is that we allow the source to retransmit old samples following erasures without resampling fresh ones. Specifically, we study the following question here:
\begin{center}
{\it how many retransmissions are to be allowed before the sample becomes too stale and useless?}
\end{center}
We carefully provide an answer to that question that depends on the privacy leakage budget and the channel statistics such that the long-term average utility is maximized.

{\bf Related works.} A number of works in the literature study the relationship between AoI and privacy. Our previous work \cite{banawan2021timely} considers an information-theoretic private information retrieval problem with AoI guarantees. The work in \cite{zhang2022age} studies differential privacy metrics that depend on AoI. Reference \cite{sathyavageesran2022privacy} is closely-related to our work. It considers the privacy-AoI tradeoff in discrete-time systems, and designs post-sample waiting policies for when to release updates in queuing systems in order to control the privacy leakage. Different from \cite{sathyavageesran2022privacy}, we consider a continuous-time system, with erasures, and jointly design waiting times and the number of retransmissions to balance AoI with leakage.

\section{System Model and Objective}

Consider a stochastic process $\{X_t\}$ that represents a time-varying status to be conveyed to a destination. Such process represents a user's status over time, e.g., home electric usage. Samples from this process are {\it generated at will,} and are sent through a channel that introduces random delays and erasures. Specifically, the $j$th sample is generated at time $S_j$, transmitted for the first time at $T_{j,1}$, and takes $b_{j,1}$ time units to traverse through the channel, denoted the channel {\it busy time}. After that, the sample is still prune to {\it erasure} with probability $\epsilon$, whence the sample may be retransmitted at time $T_{j,2}$, incurring $b_{j,2}$ channel busy time, and the process repeats. In general, the $j$th sample may be (re)transmitted $k_j$ times until successful reception. In case the $k_j$th attempt fails, the sample is discarded and the process restarts with a {\it fresh} sample $j+1$. Observe that $k_j=1$ means that the sample is transmitted only once. We now have the following constraints:
\begin{align}
T_{j,1}&\geq S_j,\quad \forall j, \label{eq_smpl-cnstrnt-1} \\
T_{j,k+1}&\geq T_{j,k}+b_{j,k},\quad \forall j,~1\leq k\leq k_j, \label{eq_smpl-cnstrnt-2}\\
S_{j+1}&\geq T_{j,k_j}+b_{j,k_j},\quad \forall j,~k_j. \label{eq_smpl-cnstrnt-3}
\end{align}
Channel busy times, $b_{j,k}$'s, are independent and identically distributed (i.i.d.). Similarly, erasure events are i.i.d., and are independent from $\{X_t\}$ and $b_{j,k}$'s.

\begin{figure}[t]
\centering
\includegraphics[scale=.45]{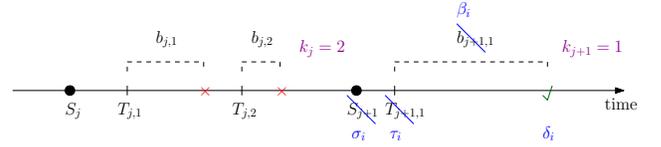}
\caption{An example time line evolution; red crosses denote failed transmissions, and the green checkmark denotes a successful transmission.}
\label{fig_time_line}
\vspace{-.2in}
\end{figure}

A successfully-received sample is denoted an {\it update}. Let $\sigma_i$, $\tau_i$ and $\beta_i$ denote the sampling time of the $i$th update, its transmission time, and the channel busy time it encounters, respectively. It follows that $\left\{\sigma_i\right\}\subseteq\{S_j\}$, $\left\{\tau_i\right\}\subseteq\{T_{j,k}\}$ and $\left\{\beta_i\right\}\subseteq\{b_{j,k}\}$. The $i$th update is delivered at time
\begin{align}
\delta_i=\tau_i+\beta_i.
\end{align}
See Fig.~\ref{fig_time_line} for an example time line including all the variables introduced so far. At the destination, the age-of-information (AoI) of the process $\{X_t\}$ at time $t$ is defined as
\begin{align}
a(t)=t-\max\left\{\sigma_i:~\delta_i\leq t\right\}.
\end{align}
We measure the utility of the status updates through a general increasing age-penalty functional $g(\cdot)$ that acts upon the AoI process $a(t)$. Specifically, the instantaneous utility of the updates at time $t$ is given by
\begin{align}
-g\left(a(t)\right).
\end{align}
Therefore, updates are more useful when their AoI is small. Observe that the {\it AoI drops right after delivery times.} We note that measuring utility through AoI is meaningful in estimation and tracking settings, as one can show that the minimum mean square error estimate of Markovian processes is given by an increasing function of the AoI, see, e.g., \cite{sun-wiener, arafa-banawan-seddik-sample}. 

Correlated with $\{X_t\}$ is another stochastic process $\{Y_t\}$ that represents a latent variable that needs to be kept {\it private} from the destination. We consider an {\it honest-but-curious} destination node that may be interested in getting more information about the user from the updates it conveys. The privacy {\it leakage} at time $t$ is governed by the amount of information that the received samples, so far, can reveal about $Y_t$, which we capture using the following non-negative function $\rho(\cdot:\cdot)$: 
\begin{align}
\rho\left(\{X_{\sigma_i}\}_{\delta_i\leq t}:Y_t\right),
\end{align}
where $\{X_{\sigma_i}\}_{\delta_i\leq t}$ denotes all the updates received up to time $t$. For instance, one can adopt the mutual information \cite{cover} to measure the privacy leakage, as done in several works \cite{sankar-tandon-poor-privacy,fawaz-privacy,medard-privacy-funnel,asoodeh-privacy,wang-ying-zhang-privacy,sankar-tan-mutual-info-privacy,khisti-privacy,gunduz-privacy}, or other notions such as $\alpha-Leakage$ in \cite{sankar-alpha-leakage,kamatsuka-voi}, and its generalization, $g-Leakage$ in \cite{sankar-g-leakage}. We have the following assumption about $\rho$:
\begin{align} \label{eq_rho-assum}
&\rho\left(\{X_{\sigma_i}\}_{\delta_i\leq t_1}:Y_{t_1}\right)\geq\rho\left(\{X_{\sigma_i}\}_{\delta_i\leq t_2}:Y_{t_2}\right), \nonumber \\
&\hspace{.75in}\forall t_1<t_2,~\text{s.t. }\left|\{\delta_i\leq t_1\}\right|=\left|\{\delta_i\leq t_2\}\right|,
\end{align}
where $|\cdot|$ denotes cardinality. Thus, the leakage decreases over time, as long as no new samples have been received. This also implies that {\it leakage peaks occur right after delivery times.} Several situations satisfy the privacy leakage assumption in (\ref{eq_rho-assum}). For instance, consider the information leakage metric to an estimating adversary in \cite{asoodeh-diaz-estimation-privacy}, 
\begin{align}
    \mathcal{L}(Y_t \rightarrow X_t)=\frac{\text{Var}[Y_t]}{\mathbb{E}[(Y_t-\mathbb{E}[Y_t|X_t])^2]},
\end{align}
where the leakage $\mathcal{L}(Y_t \rightarrow X_t)$ signifies the estimating accuracy of the adversary (i.e., $\mathcal{L}(Y_t \rightarrow X_t) \rightarrow \infty$ denotes almost perfect estimation of $Y_t$ given $X_t$). Considering the estimation setting $X_t=Y_t+W_t$, where $W_t$ is a Wiener process with $W_0=0$ and $Y_t \sim \mathcal{N}(0,\sigma_0^2)$ i.i.d. Gaussian process, then $\mathbb{E}[Y_t|X_t]=\frac{\sigma_0^2}{\sigma_0^2+t} X_t$,. The estimation leakage is given by:
\begin{align}
    \rho\left(\{X_{\sigma_i}\}_{\delta_i\leq t}:Y_t\right)=\mathcal{L}(Y_t \rightarrow X_t)=\frac{1}{1-\frac{\sigma_0^2}{\sigma_0^2+t}}
\end{align}
Hence, $ \mathcal{L}(Y_t \rightarrow X_t)$ decreases over time. The same arguments hold for the guessing adversary if $Y_t$ is picked from a discrete distribution with $\mathcal{L}(Y_t \rightarrow X_t)=\frac{\mathbb{E}[\max_{y \in \mathcal{Y}} \: P(Y_t|X_t)]}{\max_{y \in \mathcal{Y}} \: P(Y_t)}$. A different example with an Ornstein-Uhlenbeck (OU) process estimation and a mutual information leakage metric can be found in Section~\ref{sec_probForm}. 

From the above, we see a tension between utility and privacy, as noted in previous works \cite{sankar-tandon-poor-privacy,fawaz-privacy,medard-privacy-funnel,asoodeh-privacy, sankar-tan-mutual-info-privacy,gunduz-privacy,kamatsuka-voi,sankar-soheil-mohajer-privacy-utility}: {\it reducing AoI increases the leakage, and vice versa.}


Our goal is to maximize the utility of the updates, while preserving privacy. Specifically, we wish to design the sampling times $\{S_j\}$ and transmission times $\{T_{i,j}\}$ such that the long-term average age-penalty is minimized, subject to an upper bound on the average maximum privacy leakage:
\begin{align} \label{opt_main}
\min_{\{S_j\},\{T_{j,k}\}} \quad &\limsup_{n\rightarrow\infty} 
\frac{\sum_{i=1}^n\mathbb{E}\left[\int_{\delta_{i-1}}^{\delta_i}g\left(t-\sigma_{i-1}\right)dt\right]}{\sum_{i=1}^n\mathbb{E}\left[\delta_i-\delta_{i-1}\right]} \nonumber \\
\mbox{s.t.~~~} \quad &\mathbb{E}\left[\max_t \rho\left(\{X_{\sigma_i}\}_{\delta_i\leq t}:Y_t\right)\right]\leq\Delta \nonumber \\
&(\ref{eq_smpl-cnstrnt-1})-(\ref{eq_smpl-cnstrnt-3}),
\end{align}
where $\Delta\geq0$ denotes the {\it leakage budget}, and $\mathbb{E}\left[\cdot\right]$ denotes the expectation over the random variables involved.

For $\Delta=\infty$, problem (\ref{opt_main}) reduces to minimizing an AoI functional, and hence it would be optimal to set $T_{j,1}=S_j$ (first transmission times are the same as sampling times) and $k_j=1$ (only one transmission attempt per sample), $\forall j$. The reason behind these is obvious: there is no need to keep a fresh sample waiting idly {\it after} being generated as this will only hurt the AoI, and retransmitting an older sample is worse than discarding it and starting fresh. We note that this has been the typical scenario in most AoI literature that do not consider a sampling constraint budget.

Now for $\Delta<\infty$, it may be beneficial to set $T_{j,1}>S_j$. This would relatively increase the AoI of the received update, but at the same time would decrease the privacy leakage. The reason follows from the assumption in (\ref{eq_rho-assum}); the peak leakage occurs at delivery times, and we need to make it no larger than $\Delta$, on average. Following the same rationale, it may also be beneficial to retransmit the same sample in case of a failure since a fresher sample would lead to a relatively lower AoI, and hence a higher privacy leakage, when delivered.

We characterize the solution of (\ref{opt_main}) in the remainder of this paper. In particular, we discuss when to acquire a new sample, when to transmit it, and how many times it should be retransmitted so as to keep the information about $\{X_t\}$ fresh and that about $\{Y_t\}$ private.

\section{Problem Reformulation:\\Waiting Times and Stationary Policies} \label{sec_probForm}

In this section, we focus on processes in which the privacy leakage reduces, with a slight abuse of notation, to
\begin{align}
\rho\left(\{X_{\sigma_i}\}_{\delta_i\leq t}:Y_t\right)&\equiv\rho\left(t-\max\left\{\sigma_i:~\delta_i\leq t\right\}\right) \nonumber \\
&=\rho\left(a(t)\right).
\end{align}
Thus, according to our assumption on $\rho$, the privacy leakage is a decreasing function of AoI. For instance, consider an OU process $\{X_t\}$ with parameters $\sigma^2$ and $\theta$ \cite{ou-brownian-motion}, and set $Y_t=X_t+N_t$ with $N_t\sim\mathcal{N}\left(0,\sigma_0^2\right)$ being i.i.d. noise. Further, let the leakage function $\rho$ be given by the mutual information. One can show that \cite{cover}
\begin{align}
&\rho\left(\{X_{\sigma_i}\}_{\delta_i\leq t}:Y_t\right)=I\left(\{X_{\sigma_i}\}_{\delta_i\leq t};Y_t\right) 
\nonumber \\
&\hspace{.5in}=\frac{1}{2}\log\left(\frac{\frac{\sigma^2}{2\theta}+\sigma_0^2}{\frac{\sigma^2}{2\theta}\left(1-e^{-2\theta a(t)}\right)+\sigma_0^2}\right), \label{eq_rho-ou}
\end{align}
which is a decreasing function of $a(t)$ as required.

We now reformulate the optimization problem in (\ref{opt_main}) in terms of {\it waiting times,} as opposed to sampling and transmission times. Specifically, we denote by an {\it epoch} the time elapsed in between two successful updates: the $i$th epoch extends from $\delta_{i-1}$ until $\delta_i$. We now define the first {\it pre-sampling waiting time} in the $i$th epoch as
\begin{align}
W_{i,1}\triangleq \min\left\{S_j:~S_j\geq\delta_{i-1}\right\}-\delta_{i-1}.
\end{align}
That is, $W_{i,1}$ is the waiting time at the beginning of epoch $i$ before acquiring the {\it first} sample in it. Next, we define the first {\it post-sampling waiting time} following the acquisition of the first sample in the $i$th epoch as
\begin{align}\label{eq_pst-smpl-1}
Z^1_{i,1}\triangleq \min\left\{T_{j,1}:~T_{j,1}\geq\delta_{i-1}\right\}-\left(\delta_{i-1}+W_{i,1}\right).
\end{align}
We now (re)denote by $b^k_{i,1}$ the channel busy time of the first sample's $k$th transmission attempt in the $i$th epoch. Therefore, we have the post-sampling waiting time sequence given by
\begin{align}\label{eq_pst-smpl-k}
&Z^k_{i,1}\triangleq \min\left\{T_{j,1}:~T_{j,1}\geq\delta_{i-1}+\sum_{l=1}^{k-1} Z^l_{i,1}+b^l_{i,1}\right\} \nonumber \\
&-\left(\delta_{i-1}+W_{i,1}+\sum_{l=1}^{k-1} Z^l_{i,1}+b^l_{i,1}\right),\quad 1\leq k\leq \kappa_{i,1},
\end{align}
where $\kappa_{i,1}$ now denotes the maximum number of transmission attempts for the first sample in the $i$th epoch. For simplicity of presentation, let us define
\begin{align}
M_{i,1}\triangleq\delta_{i-1}+W_{i,1}+\sum_{k=1}^{\kappa_{i,1}} Z^k_{i,1}+b^k_{i,1}
\end{align}
as the maximum time allocated to the transmission attempts of the first sample in the $i$th epoch. In case the first sample is unsuccessful after its last transmission attempt, we define the {\it second} pre-sampling waiting time as
\begin{align}
W_{i,2}\triangleq\min\left\{S_j:~S_j\geq M_{i,1}\right\}-M_{i,1}.
\end{align}
This is then followed by a post-sampling waiting time sequence $\{Z^k_{i,2}\}$ given exactly as in (\ref{eq_pst-smpl-1}) and (\ref{eq_pst-smpl-k}) after replacing $\delta_{i-1}$, $W_{i,1}$ and $\kappa_{i,1}$ by $M_{i,1}$, $W_{i,2}$ and $\kappa_{i,2}$, respectively.

In general, the pre- and post-sampling waiting time sequences in the $i$th epoch will be given as follows:
\begin{align}
W_{i,r}=&\min\left\{S_j:~S_j\geq M_{i,r-1}\right\}-M_{i,r-1}, \\
M_{i,r}=&M_{i,r-1}+W_{i,r}+\sum_{k=1}^{\kappa_{i,r}} Z^k_{i,r}+b^k_{i,r}, \\
Z^k_{i,r}=&\min\left\{T_{j,1}:~T_{j,1}\geq M_{i,r-1}+\sum_{l=1}^{k-1} Z^l_{i,1}+b^l_{i,1}\right\} \nonumber \\
&\hspace{-.25in}-\left(M_{i,r-1}+W_{i,r}+\sum_{l=1}^{k-1} Z^l_{i,r}+b^l_{i,r}\right),\quad k\leq \kappa_{i,r},
\end{align}
with $r\geq1$ and $M_{i,0}\triangleq\delta_{i-1}$. The $i$th epoch ends whenever a transmission attempt is successful, which defines $\delta_i$ and the start of the next epoch $i+1$. The length of the $i$th epoch is 
\begin{align} \label{eq_L}
L_i=\delta_i-\delta_{i-1} =&\sum_{r=1}^{R_i-1}\left(W_{i,r}+\sum_{k=1}^{\kappa_{i,r}}Z^k_{i,r}+b^k_{i,r}\right) \nonumber \\
&+W_{i,R_i}+\sum_{k=1}^{\psi_i}Z^k_{i,R_i}+b^k_{i,R_i},
\end{align}
where $R_i$ is the number of samples generated in the $i$th epoch, and $\psi_i$ denotes the number of attempts needed for the $R_i$th sample to be delivered. Clearly, $\psi_i\leq\kappa_{i,R_i}$. Observe that the AoI at the start of epoch $i+1$ is given by
\begin{align}
a\left(\delta_i\right)=\sum_{k=1}^{\psi_i}Z^k_{i,R_i}+b^k_{i,R_i}.
\end{align}

We focus on {\it stationary} policies in which the waiting times have the same distribution across epochs. We also fix 
\begin{align}
\kappa_{i,r}=K,\quad\forall i,r.
\end{align}
Problem (\ref{opt_main}) now reduces to one over a single epoch:
\begin{align} \label{opt_wait}
\min_{\{W_{i,r}\geq0\},~\{Z^k_{i,r}\geq0\},~K\in\mathbb{Z}_{++}} \quad &\frac{\mathbb{E}\left[\int_0^{L_i}g\left(a\left(\delta_{i-1}\right)+t\right)dt\right]}{\mathbb{E}\left[L_i\right]} \nonumber \\
\mbox{s.t.\hspace{.6in}}\quad&\mathbb{E}\left[\rho\left(\sum_{k=1}^{\psi_i}Z^k_{i,R_i}+b^k_{i,R_i}\right)\right]\!\leq\!\Delta. 
\end{align}

We now have the following lemma:

\begin{lemma} \label{thm_wait}
In problem (\ref{opt_wait}), it is optimal to perform pre-sampling waiting only at the beginning of an epoch. Likewise, it is optimal to perform post-sampling waiting only once per sample following each sampling time.
\end{lemma}

\begin{Proof}
Observe that the epoch length in (\ref{eq_L}) only depends on the aggregate sum of the pre-sampling waiting times. One can then define the aggregate pre-sampling waiting time
\begin{align}
W_i\triangleq\sum_{r=1}^{R_i}W_{i,r},
\end{align}
and optimize that instead, which does not change the value of the optimal solution. Similarly, one can also define aggregate post-sampling waiting times
\begin{align}
Z_{i,r}\triangleq\sum_{k=1}^KZ^k_{i,r},~1\leq r\leq R_i-1,\quad Z_{i,R_i}\triangleq\sum_{k=1}^{\psi_i}Z^k_{i,r},
\end{align}
and optimize those instead.
\end{Proof}

Next, we focus on {\it deterministic waiting policies} in which the pre-sampling waiting time $W_i$ is a deterministic function of the starting AoI of the $i$th epoch:
\begin{align}
W_i\equiv w\left(Z_{i-1,R_{i-1}}+\sum_{k=1}^{\psi_{i-1}}b^k_{i-1,R_{i-1}}\right).
\end{align} 
We note that stationary deterministic waiting policies are known to be optimal under i.i.d. channel settings \cite{sun-age-mdp}. By Lemma~\ref{thm_wait}, and under stationary deterministic waiting policies, the $i$th epoch length is now given by
\begin{align} \label{eq_L-stationary}
L_i=&w\left(Z_{i-1,R_{i-1}}+\sum_{k=1}^{\psi_{i-1}}b^k_{i-1,R_{i-1}}\right)+\sum_{r=1}^{R_i}Z_{i,r} \nonumber \\
&+\sum_{r=1}^{R_i-1}\sum_{k=1}^Kb^k_{i,r}+\sum_{k=1}^{\psi_i}b^k_{i,R_i},
\end{align}
and the optimization problem finally becomes
\begin{align} \label{opt_wait-aggr}
\min_{w(\cdot)\geq0,~\{Z_{i,r}\geq0\},~K\in\mathbb{Z}_{++}} \quad &\frac{\mathbb{E}\left[\int_0^{L_i}g\left(a\left(\delta_{i-1}\right)+t\right)dt\right]}{\mathbb{E}\left[L_i\right]} \nonumber \\
\mbox{s.t.\hspace{.55in}}\quad&\mathbb{E}\left[\rho\left(Z_{i,R_i}+\sum_{k=1}^{\psi_i}b^k_{i,R_i}\right)\right]\leq\Delta. 
\end{align}

In the sequel, we present solutions to problem (\ref{opt_wait-aggr}) first for the case without errors, followed by that with errors.

\section{The Case Without Errors: $\epsilon=0$}

We analyze problem (\ref{opt_wait-aggr}) for error-free transmissions in this section. Our structural insights drawn from the solution of this scenario will serve as a building block for the scenario with channel errors in the following section. Now, for $\epsilon=0$, we have $K=1$ (no retransmissions are needed), $R_i=1,~\forall i$ and $\psi_i=1,~\forall i$. Hence, we drop the $r$ subscript and $k$ superscript in $Z_{i,r}$ and $b^k_{i,r}$, and re-evaluate the epoch length in (\ref{eq_L-stationary}) as
\begin{align} \label{eq_L-no-err}
L_i=&w\left(Z_{i-1}+b_{i-1}\right)+Z_i+b_i.
\end{align}
Problem (\ref{opt_wait-aggr}) then reduces to
\begin{align} \label{opt_no-err}
\min_{w(\cdot)\geq0,~\{Z_i\geq0\}} \quad &\frac{\mathbb{E}\left[\int_0^{L_i}g\left(a\left(\delta_{i-1}\right)+t\right)dt\right]}{\mathbb{E}\left[L_i\right]} \nonumber \\
\mbox{s.t.~~~~~~}\quad&\mathbb{E}\left[\rho\left(Z_i+b_i\right)\right]\leq\Delta. 
\end{align}

\begin{lemma} \label{thm_post-wait}
The optimal post-sampling waiting policy of problem (\ref{opt_no-err}) is $Z_i^*=\zeta,~\forall i$, for some constant $\zeta$ given by
\begin{align} \label{eq_opt-post-wait}
\zeta=\begin{cases}0,\quad&\text{if }\mathbb{E}\left[\rho\left(b_i\right)\right]<\Delta\\ \xi(\Delta),\quad&\text{otherwise}\end{cases},
\end{align}
where $\xi(\Delta)$ is the unique solution of $\mathbb{E}\left[\rho\left(\xi(\Delta)+b_i\right)\right]=\Delta$.
\end{lemma}

\begin{Proof}
We first argue that $Z_i$ cannot depend on $Z_j,~j\leq i-1$, since the new sample generated in the $i$th epoch leaks information at the beginning of epoch $i+1$ with a value that is independent of previous epochs' events. Specifically, $Z_i$ depends solely on $b_i$ (through its distribution). Since $b_i$'s are i.i.d., we can conclude that $Z_i$'s are also i.i.d.

Next, observe that the post-sampling waiting time $Z_i$ can only hurt the AoI. This can readily be shown by a sample path argument; increasing $Z_i$ increases the service time of the $i$th sample, and only makes it more stale when received. Now let us fix $Z_{i-1}$. Setting $Z_i=0$ would then be AoI-optimal, provided that the privacy constraint is met, i.e., if $\mathbb{E}\left[\rho\left(b_i\right)\right]<\Delta$. Otherwise, one should set $Z_i$ to the lowest value allowed by the leakage budget. Since $\rho$ is decreasing, such value is given by $\xi(\Delta)$ in (\ref{eq_opt-post-wait}).

Finally, since $Z_i$'s are i.i.d., the above argument shows that they should all be fixed at the same value.
\end{Proof}

Lemma~\ref{thm_post-wait} shows that there can be situations in which the channel busy time provides a {\it natural} privacy cover (when $\mathbb{E}\left[\rho\left(b_i\right)\right]<\Delta$), and that post-waiting times should only be used when necessary. The lemma also shows that one can define a {\it new} channel busy time
\begin{align}
\tilde{b}_i\triangleq b_i+\zeta,\quad \forall i,
\end{align}
with $\zeta$ given by (\ref{eq_opt-post-wait}), which is still i.i.d., and optimize the pre-sampling waiting time over $\{\tilde{b}_i\}$ as done in the AoI minimization literature. Specifically, the results in, e.g., \cite{sun-cyr-aoi-non-linear,arafa-banawan-seddik-sample}, show that the pre-sampling policy is a {\it threshold policy}
\begin{align}
w\left(t\right)=\left[G^{-1}_t\left(\gamma\right)\right]^+,
\end{align}
where $\left[\cdot\right]^+\triangleq\max\left(\cdot,0\right)$, and the function
\begin{align}
G_t\left(x\right)\triangleq\mathbb{E}\left[g\left(t+x+\tilde{b}_i\right)\right]
\end{align}
denotes the expected utility by the end of the $i$th epoch when it starts with an AoI value of $t$. Further, the value of $\gamma$ is given by the unique solution of 
\begin{align}
&\mathbb{E}\left[\int_0^{\left[G^{-1}_{\tilde{b}_{i-1}}\left(\gamma\right)\right]^++\tilde{b}_i}g\left(\tilde{b}_{i-1}+t\right)dt\right] \nonumber \\
&\hspace{1in}-\gamma\mathbb{E}\left[\left[G^{-1}_{\tilde{b}_{i-1}}\left(\gamma\right)\right]^++\tilde{b}_i\right]=0,
\end{align}
which can be found by, e.g., a bisection search \cite{arafa-banawan-seddik-sample}.

This completes the solution for the setting without errors.

\section{The Case With Errors: $\epsilon>0$}

In this section, we extend the aforementioned solution to the case with channel errors. First, we fix the value of $K$ and evaluate the distributions of the random variables $R_i$ and $\psi_i$. Observe that a new sample will be generated only if the previous one has $K$ failed transmissions, which occurs with probability $\epsilon^K$. It then follows that
\begin{align} \label{eq_Ri-distrb}
R_i\sim\text{geometric}\left(1-\epsilon^K\right),\quad\forall i.
\end{align}
As for the number of transmissions, $\psi_i$, needed for sample $R_i$ (the final sample) to succeed, we note that $\psi_i=k$ in case the $k$th transmission attempt is successful given that a successful transmission occurs in at most $K$ attempts. Thus,
\begin{align} \label{eq_psi-distrb}
\mathbb{P}\left(\psi_i=k\right)=\frac{\epsilon^{k-1}(1-\epsilon)}{1-\epsilon^K},~1\leq k\leq K,\quad \forall i,
\end{align}
i.e., $\psi_i$ is a truncated $\sim\text{geometric}(1-\epsilon)$ random variable.

Next, following similar arguments as in Lemma~\ref{thm_post-wait}, one can show that the post-sampling waiting times $Z_{i,r}$'s are all fixed in the optimal solution of problem (\ref{opt_wait-aggr}), and are given by (\ref{eq_opt-post-wait}). Hence, the epoch length in (\ref{eq_L-stationary}) is now proportional to
\begin{align}
R_i\zeta.
\end{align}
This allows us to draw the following insight:

\begin{remark}
As the privacy leakage budget $\Delta$ decreases, $\zeta$ increases, and hence the value of $R_i$ must be relatively small so as to not to make the epoch length too large. This can be achieved by increasing $K$, which controls the distribution of $R_i$ and makes it take smaller values with higher probabilities. 
\end{remark}

The above remark is one fundamental observation in this paper: \emph{the number of retransmissions $K$ should be inversely proportional to $\Delta$.} Intuitively, higher levels of privacy are naturally achieved when retransmitting stale samples, and therefore making a case for retransmission without resampling.

To get more insight on how the utility behaves as a function of $K$, we focus on the scenario in which $g(x)=x$, together with a zero-pre-sampling waiting policy in which $w(t)=0$. In this case, the starting AoI in the $i$th epoch is given by
\begin{align}
\zeta+\sum_{k=1}^{\psi_{i-1}}b^k_{i-1,R_{i-1}},
\end{align}
and the $i$th epoch length in (\ref{eq_L-stationary}) reduces to
\begin{align}
L_i=&R_i\zeta+\sum_{r=1}^{R_i-1}\sum_{k=1}^Kb^k_{i,r}+\sum_{k=1}^{\psi_i}b^k_{i,R_i}.
\end{align}
Consequently, direct geometrical arguments lead to expressing the utility (long-term average AoI) as
\begin{align} \label{eq_aoi-smpl}
\zeta+&\mathbb{E}\left[\psi_i\right]\mathbb{E}\left[b_i\right]\!+\!\frac{\frac{1}{2}\mathbb{E}\left[L_i^2\right]}{\mathbb{E}\left[R_i\right]\zeta+\mathbb{E}\left[R_i\!-\!1\right]K\mathbb{E}\left[b_i\right]+\mathbb{E}\left[\psi_i\right]\mathbb{E}\left[b_i\right]}. 
\end{align}
\vspace{-.2in}

We start with computing the second moment of $L_i$. Since $\left\{b^k_{i,r}\right\}$, $R_i$ and $\psi_i$ are mutually independent, one can write
\begin{align} \label{eq_L2-0}
\mathbb{E}\left[L_i^2\right]=&\mathbb{E}\left[\left(R_i\zeta+\sum_{r=1}^{R_i-1}\sum_{k=1}^Kb^k_{i,r}\right)^2\right]+\mathbb{E}\left[\left(\sum_{k=1}^{\psi_i}b^k_{i,R_i}\right)^2\right] \nonumber \\
&+2\mathbb{E}\left[R_i\zeta+\sum_{r=1}^{R_i-1}\sum_{k=1}^Kb^k_{i,r}\right]\mathbb{E}\left[\sum_{k=1}^{\psi_i}b^k_{i,R_i}\right]. 
\end{align}
One can then show that the second term in (\ref{eq_L2-0}) is given by
\begin{align} \label{eq_L2-1}
\mathbb{E}\left[\left(\sum_{k=1}^{\psi_i}b^k_{i,R_i}\right)^2\right]\!=\!\mathbb{E}\left[\psi_i\right]\mathrm{Var}\left[b_i\right]+\mathbb{E}\left[\psi_i^2\right]\left(\mathbb{E}\left[b_i\right]\right)^2,
\end{align}
while the first term in (\ref{eq_L2-0}) can be expressed as
\begin{align} \label{eq_L2-2}
\mathbb{E}&\left[\left(R_i\zeta+\sum_{r=1}^{R_i-1}\sum_{k=1}^Kb^k_{i,r}\right)^2\right]=\mathbb{E}\left[R_i^2\right]\zeta^2 \nonumber \\
&\!+\!\mathbb{E}\!\left[\!\left(\sum_{r=1}^{R_i-1}\sum_{k=1}^Kb^k_{i,r}\right)^{\!\!2}\right]\!\!+\!2\!\left(\mathbb{E}\!\left[R_i^2\right]\!-\!\mathbb{E}\left[R_i\right]\right)\!\zeta K\mathbb{E}\left[b_i\right],
\end{align}
where the last term follows by iterated expectations. Focusing on the second term above, one can expand it as follows:
\begin{align} \label{eq_L2-3}
\mathbb{E}\left[\left(\sum_{r=1}^{R_i-1}\sum_{k=1}^Kb^k_{i,r}\right)^2\right]=&\mathbb{E}\left[\sum_{r=1}^{R_i-1}\left(\sum_{k=1}^Kb^k_{i,r}\right)^2\right] \nonumber \\
&\hspace{-.5in}+\mathbb{E}\left[\sum_{\substack{r,r^\prime=1 \\ r^\prime\neq r}}^{R_i-1}\left(\sum_{k=1}^Kb^k_{i,r}\right)\left(\sum_{k^\prime=1}^Kb^{k^\prime}_{i,r^\prime}\right)\right] \\
&\hspace{-1.4in}=\mathbb{E}\left[R_i-1\right]\left(K\mathbb{E}\left[b_i^2\right]+K(K-1)\left(\mathbb{E}\left[b_i\right]\right)^2\right) \nonumber \\
&\hspace{-.5in}+\mathbb{E}\left[R_i-1\right]\mathbb{E}\left[R_i-2\right]K^2\left(\mathbb{E}\left[b_i\right]\right)^2 \\
&\hspace{-1.4in}=\mathbb{E}\left[R_i-1\right]K\mathrm{Var}\left[b_i\right]+\left(\mathbb{E}\left[R_i-1\right]\right)^2K^2\left(\mathbb{E}\left[b_i\right]\right)^2. \label{eq_L2-3}
\end{align}
Substituting (\ref{eq_L2-3}) in (\ref{eq_L2-2}), and then (\ref{eq_L2-2}) and (\ref{eq_L2-1}) in (\ref{eq_L2-0}), we get an expression for the second moment of $L_i$ in terms of the first and second moments of $R_i$ and $\psi_i$. These latter moments are directly computable from (\ref{eq_Ri-distrb}) and (\ref{eq_psi-distrb}) as
\begin{align}
\mathbb{E}\left[R_i\right]=&\frac{1}{1-\epsilon^K},~\mathbb{E}\left[R_i^2\right]=\frac{1+\epsilon^K}{\left(1-\epsilon^K\right)^2} \\
\mathbb{E}\left[\psi_i\right]=&\frac{1-\left(K+1\right)\epsilon^K+K\epsilon^{K+1}}{\left(1-\epsilon^K\right)\left(1-\epsilon\right)}, \nonumber \\
\mathbb{E}\left[\psi_i^2\right]=&\frac{1}{\left(1-\epsilon^K\right)\left(1-\epsilon\right)^2}\left[1+\epsilon-\left(K+1\right)^2\epsilon^K\right. \nonumber \\
&\qquad\left.+\left(2K^2+2K-1\right)\epsilon^{K+1}-K^2\epsilon^{K+2}\right].
\end{align}

Finally, observe that the remaining terms in (\ref{eq_aoi-smpl}) are merely the first moments of $R_i$ and $\psi_i$ computed above. We now have a closed-form expression of the utility in terms of the number of retransmissions $K$, and the privacy leakage budget $\Delta$ (embedded in the value of $\zeta$).

Next, we discuss how to choose the optimal $K$.

\section{Optimal Number of Retransmissions}

\begin{figure}[t]
\centering
\includegraphics[scale=.45]{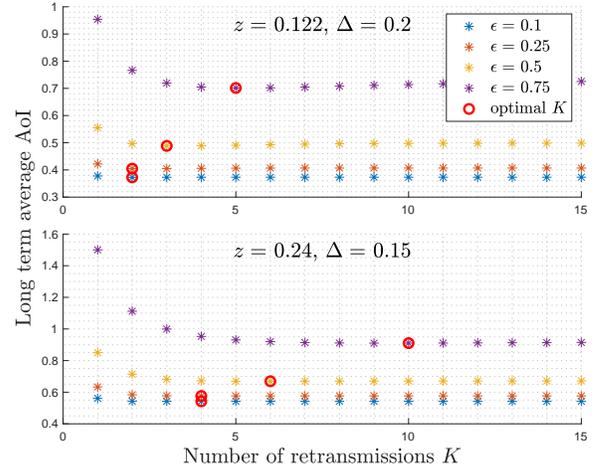}
\caption{Utility vs.~retransmissions; channel busy time $\sim\exp(10)$.}
\label{fig_lmda10}
\vspace{-.2in}
\end{figure}

We evaluate the optimal number of retransmissions $K^*$ for a system with $b_i\sim\exp(\lambda)$. For the privacy leakage function, we consider the OU process example considered in Section~\ref{sec_probForm} ($\rho$ is given by (\ref{eq_rho-ou})). In Fig.~\ref{fig_lmda10}, we plot the utility versus $K$ for different values of error probability $\epsilon$. In the top figure, we consider a leakage budget of $\Delta=0.2$, for which one can show by (\ref{eq_opt-post-wait}) that $\zeta=0.122$. In the bottom figure, we consider $\Delta=0.15$, for which $\zeta=0.24$. In both cases, $\lambda=10$. Evidently, {\it the utility in the bottom figure is worse (higher AoI) since the leakage budget is tighter.} Two further observations can be drawn. First, the optimal $K^*$ (circled in red) increases with $\epsilon$ for fixed $\zeta$. The intuition behind this is that as $\epsilon$ increases, a sample takes a relatively longer time to be successfully delivered. Hence, if one resamples often in this case (i.e., if $K$ is small), the sample will incur even more time due to the extra post-sampling waiting $\zeta$ that has to be added. The second observation is that the optimal $K^*$ increases with $\zeta$ for fixed $\epsilon$. This is also intuitive since for larger $\zeta$ the leakage budget is tighter, and hence one has to retransmit the same sample for longer times to preserve privacy.

Next, we fix $\Delta=0.2$ (i.e., $\zeta=0.122$) and vary the busy time statistic, $\lambda$. For $\lambda=1$ (slower channel) we get $K^*=1$ for all values of $\epsilon$ in Fig.~\ref{fig_lmda10}. While for $\lambda=20$ (faster channel), we get $K^*=[4,4,6,10]$, i.e., $K^*$ increases relative to the values in Fig.~\ref{fig_lmda10}. This is mainly because the channel naturally adds a privacy coverage when it is slow, and requires more protection by retranmissions when it is fast.

\section{Conclusion}

A freshness-privacy tradeoff has been considered for a source-destination pair communicating through an erasure channel. It has been shown that carefully designing {\it post-sampling waiting times,} together with the {\it number of retransmissions of unsuccessful samples} can provide useful status updates while maintaining desired levels of privacy.

\end{document}